\documentclass[conference]{IEEEtran}
\IEEEoverridecommandlockouts

\usepackage{cite}
\usepackage{amsmath,amssymb,amsfonts}
\usepackage{algorithmic}
\usepackage{graphicx}
\usepackage{textcomp}
\usepackage{xcolor}
\usepackage{hyperref}
\usepackage{booktabs}
\usepackage{multirow}
\usepackage{float}
\usepackage{url}
\usepackage{glossaries}
\usepackage[utf8]{inputenc}
\usepackage[T1]{fontenc}
\usepackage[table]{xcolor}

\makeglossaries

\newacronym{aes}{AES}{Advanced Encryption Standard}
\newacronym{aw}{AW}{Activity Watch}
\newacronym{c2}{C2}{Command and Control}
\newacronym{dll}{DLL}{Dynamic-Link Library}
\newacronym{dns}{DNS}{Domain Name System}
\newacronym{elk}{ELK}{Elasticsearch, Logstash, and Kibana}
\newacronym{ftp}{FTP}{File Transfer Protocol}
\newacronym{ip}{IP}{Internet Protocol}
\newacronym{json}{JSON}{JavaScript Object Notation}
\newacronym{llm}{LLM}{Large Language Model}
\newacronym{lora}{LoRA}{Low-Rank Adaptation}
\newacronym{mitre-attack}{ATT\&CK}{MITRE Adversarial Tactics, Techniques, and Common Knowledge}
\newacronym{nat}{NAT}{Network Address Translation}
\newacronym{pcap}{PCAP}{Packet Capture}
\newacronym{pdf}{PDF}{Portable Document Format}
\newacronym{pii}{PII}{Personally Identifiable Information}
\newacronym{rat}{RAT}{Remote Access Trojan}
\newacronym{slm}{SLM}{Small Language Model}
\newacronym{smb}{SMB}{Server Message Block}
\newacronym{sysmon}{Sysmon}{System Monitor}
\newacronym{url}{URL}{Uniform Resource Locator}
\newacronym{usb}{USB}{Universal Serial Bus}
\newacronym{vm}{VM}{Virtual Machine}
\newacronym{wbpv}{WBPV}{WebBrowserPassView}
\newacronym{wel}{WEL}{Windows Event Log}
\newacronym{wts}{WTS}{Windows Task Scheduler}

\def\BibTeX{{\rm B\kern-.05em{\sc i\kern-.025em b}\kern-.08em
    T\kern-.1667em\lower.7ex\hbox{E}\kern-.125emX}}
\begin{document}

\title{Multi-Source Cybersecurity Logs: An ATT\&CK-Labeled Dataset and SLM Evaluation}

\author{
    \IEEEauthorblockN{
        Abir Ashab Niloy\IEEEauthorrefmark{1},
        Ahmed Ryan\IEEEauthorrefmark{2},
        Imamul Hossain Rafi\IEEEauthorrefmark{1},
        Md Erfan\IEEEauthorrefmark{2}, and
        Md Rayhanur Rahman\IEEEauthorrefmark{2}
    }
    \IEEEauthorblockA{\IEEEauthorrefmark{1}University of Dhaka, Email: \{bsse1315, bsse1323\}@iit.du.ac.bd}
    \IEEEauthorblockA{\IEEEauthorrefmark{2}The University of Alabama, Email: \{aryan9, merfan\}@crimson.ua.edu, mrahman87@ua.edu}
}

\maketitle
\begin{abstract}

Multi-stage cyberattacks span system, network, and browser logs. 
Detecting them requires correlating events across all three sources. 
Machine learning methods can learn these cross-source patterns, but they need labeled multi-source data.
Existing public datasets fall short. 
Network-only datasets such as CICIDS and UNSW-NB15 miss host and browser activity. 
Host-focused datasets such as LMDG and CICAPT-IIoT lack browser telemetry. 
ATLAS includes all three sources but labels events only as malicious or benign, without \gls{mitre-attack} technique granularity.
No public dataset combines all three sources with per-entry \gls{mitre-attack} technique labels.
We close the gap by building a multi-source log dataset of 870 sessions (70 attack, 800 benign) and approximately 2.3 million events. 
We captured system, network, and browser activity simultaneously on Windows endpoints. 
We labeled malicious events with \gls{mitre-attack} technique IDs, covering 12 tactics and 53 techniques. 
We generated all attack data using real tools, including \gls{rat}, \gls{c2} tunnels, and cloud exfiltration.
To demonstrate learnability, we fine-tuned three \glspl{slm} (Qwen2.5-1.5B, Llama-3.2-3B, Phi-4-Mini) using \gls{lora}. 
We compared each against its base variant across ten metrics on two tasks: chunk classification and \gls{mitre-attack} technique identification. 
Fine-tuning improved every model on every metric. 
Chunk classification accuracy rose from approximately 8\% in the base variants to between 90\% and 97\% after fine-tuning.
Technique identification remained challenging, with the best exact-match accuracy at 42\%, although high partial-match scores show the models captured most of the underlying reasoning.

\end{abstract}


\section{Introduction}

Modern organizations record system, network, and browser activity as logs. 
Security teams rely on these logs to detect attacks before damage occurs. 
Adversaries execute multi-stage attacks that traverse all three sources~\cite{hutchins2011killchain}. 
They phish a user through a browser, run code on the system, and beacon over the network.
Reconstructing the full attack requires correlating events across all sources. 
We define the key terms (multi-source log, event, session, chunk, \gls{mitre-attack} tactic and technique~\cite{strom2020attack}) in Section~\ref{sec:key-concepts}.

When defenders see only one source, the attacker stays hidden in the others.
Initial access escalates to data theft, ransomware, or system wipeout~\cite{cisa2024ransomhub}.
82\% of threat detections in 2025 involved malware-free tactics~\cite{crowdstrike2026globalthreat}. 
Defenders cannot rely on file signatures alone. 
They must link logs across sources to uncover the attacker's true intent.
The challenge is two-fold. 
First, machine learning detection methods need labeled multi-source data to learn cross-source patterns. 
Second, no such labeled dataset is publicly available.

We address the problem by building a labeled multi-source log dataset. 
The dataset captures system, network, and browser events simultaneously. 
It tags malicious events with \gls{mitre-attack} technique IDs. 
We generated all attack data using real attack tools, not synthetic traffic generators.
However, several existing datasets already cover parts of the problem. 
KDD Cup 1999, NSL-KDD~\cite{tavallaee2009nslkdd}, CICIDS~\cite{sharafaldin2018cicids}, UNSW-NB15~\cite{moustafa2015unswnb15}, and CTU-13~\cite{garcia2014ctu13} record only network traffic and lack host and browser visibility. 
DAPT 2020~\cite{myneni2020dapt}, Unraveled~\cite{myneni2023unraveled}, CICAPT-IIoT~\cite{ghiasvand2024cicaptiiot}, and LMDG~\cite{mabrouk2025lmdg} include host telemetry but omit browser activity. 
The closest prior work, ATLAS~\cite{alsaheel2021atlas} and its successor ATLASv2~\cite{riddle2024atlasv2}, include browser logs alongside system audit and DNS data, but label events only as malicious or benign without \gls{mitre-attack} technique granularity. 
UWF-ZeekData22~\cite{bagui2023uwfzeekdata22} applies \gls{mitre-attack} labels to Zeek network logs but lacks host and browser sources, and the data is dominated by reconnaissance activity (99.97\%).

No public dataset combines all three properties at once: (a) simultaneous system, network, and browser sources, (b) per-entry \gls{mitre-attack} technique labels on malicious events, and (c) generation from real attack tooling. 
This gap prevents researchers from training and evaluating models that reason across all three sources at the technique level. \textit{The goal of this paper is to enable research on multi-source attack detection by releasing a multi-source log dataset with per-entry \gls{mitre-attack} technique labels and benchmarking fine-tuned \glspl{slm}~\cite{wang2024slmsurvey} on it, giving security and ML researchers a foundation for training cross-source detection models and giving SOC analysts an evaluated baseline for initial triage.} We investigate the following research questions.

$RQ_1$: Can attack simulations produce a multi-source log dataset with per-entry \gls{mitre-attack} technique labels and broad tactic coverage across system, network, and browser sources?

$RQ_2$: Does the resulting dataset carry a learnable signal such that \gls{lora} fine-tuning improves \glspl{slm} over their base variants on attack detection and technique identification tasks?

To answer the questions, we ran 70 attack simulations and 800 benign sessions on Windows endpoints from January 2025 to February 2026. 
Each session captured 20 minutes of system, network, and browser activity simultaneously. 
We labeled every session as normal or suspicious and tagged malicious events with \gls{mitre-attack} technique IDs. 
We then fine-tuned three \glspl{slm} (Qwen2.5-1.5B, Llama-3.2-3B, and Phi-4-Mini) using LoRA, and compared each base model against its fine-tuned variant across ten metrics.

(a) A multi-source log dataset of 870 sessions (70 attack sessions and 800 benign sessions), built from simulations with genuine attack tools, that records about 2.3 million events from system, network, and browser sources, with coverage of 12 \gls{mitre-attack} tactics and 53 techniques.
(b) A benchmark evaluation of three \glspl{slm} in base and \gls{lora}-fine-tuned form, measured across ten metrics on two tasks: chunk classification (normal vs. suspicious) and \gls{mitre-attack} technique identification.
(c) An empirical baseline showing that fine-tuning lifts chunk-classification accuracy from approximately 8\% to between 90\% and 97\%, providing direct evidence that the dataset carries learnable signal.

The remainder of this paper is organized as follows. 
Section~\ref{sec:key-concepts} defines the key concepts. 
Section~\ref{sec:related-work} reviews the related work. 
Section~\ref{sec:methodology} presents the methodology. 
Section~\ref{sec:findings} reports the findings for both research questions.
Section~\ref{discussion} discusses the implications.
Section~\ref{sec:threats-to-validity} covers the threats to validity. 
Section \ref{sec:conclusion} concludes.

\section{Key Concepts}
\label{sec:key-concepts}

We discuss the key concepts in this section.

\textit{Multi-Source Log:} 
\label{subsec:multi-source-log}
A multi-source log records activity from three sources simultaneously: system, network, and browser. 
System logs capture operating-system events; network logs capture all traffic events; and browser logs capture visited \glspl{url} and browser events. 

\textit{Event:}
\label{subsec:event}
An event is a single recorded activity from one source, such as a process starting, a network connection, or a visited \gls{url}. 

\textit{Session:}
\label{subsec:session}
A session is a fixed 20-minute window of recorded activity that consists of many events.

\textit{Chunk:}
\label{subsec:chunk}
A chunk is a structured metadata object containing an ordered sequence of seven timestamped events.
This unit serves as the input for \glspl{slm}, which process the sequence to perform classification and identify \gls{mitre-attack} techniques.

\begin{table}[h!]
\vspace{-10pt}
\scriptsize
\centering
\caption{A Chunk (Comprises Seven Events)}
\begin{tabular}{|c|l|l|l|}
\hline
\textbf{\#} & \textbf{Timestamp} & \textbf{Event} & \textbf{Technique ID} \\ \hline
1 & 09:19:00 & Sysmon 1 — powershell.exe (encoded cmd) & T1059.001 \\ \hline
2 & 09:19:15 & PCAP — 192.168.1.100 $\to$ 147.185.221.22 & T1071.001 \\ \hline
3 & 09:19:30 & Sysmon 3 — powershell.exe network conn & T1071.001 \\ \hline
4 & 09:19:45 & Sysmon 1 — explorer.exe (normal startup) & -- \\ \hline
5 & 09:20:00 & Browser — HTTPS to accounts.google.com & -- \\ \hline
6 & 09:20:15 & Sysmon 4688 — notepad.exe & -- \\ \hline
7 & 09:21:30 & Sysmon 11 — .txt.encrypted file created & T1486 \\ \hline
\end{tabular}
\\[3pt]
\footnotesize \textit{Note:} Each event is either a system, network or browser log. A chunk is labeled as suspicious if associated \gls{mitre-attack} technique sum $\ge1$.
\vspace{-5pt}
\end{table}

\textit{MITRE ATT\&CK Tactics and Techniques:}
\label{subsec:mitre-tactics-techniques}
\gls{mitre-attack} is a public knowledge base of adversary behavior, in which a tactic is the attacker's goal, such as initial access or exfiltration, and a technique is the specific method to reach the goal \cite{strom2020attack}. 

\textit{Small Language Model (SLM):} 
\label{subsec:slm}
A small language model is a language model with less than 10 billion parameter count typically~\cite{wang2024slmsurvey}.

\textit{Fine Tuning:}
\label{subsec:fine-tuning}
Fine-tuning is the process of training a pretrained model further on task-specific data to adapt the model's behavior. 

\textit{Low-Rank Adaptation (LoRA):}
\label{subsec:lora}
\gls{lora} is a fine-tuning method that freezes the original model weights and trains only a small set of added adapter layers \cite{hu2022lora}. 

\section{Related Work}
\label{sec:related-work}

We discuss related works in this section.

\textit{Intrusion-detection datasets:} 
Early datasets are network-only and miss host and browser activity. 
KDD Cup 1999 and NSL-KDD~\cite{tavallaee2009nslkdd} are over 25 years old and lack modern threats. 
CICIDS~\cite{sharafaldin2018cicids}, UNSW-NB15~\cite{moustafa2015unswnb15}, and CTU-13~\cite{garcia2014ctu13} record only network traffic. 
UWF-ZeekData22~\cite{bagui2023uwfzeekdata22} adds \gls{mitre-attack} labels but is reconnaissance-dominated (99.97\%) and network-only. 
The DARPA Transparent Computing dataset~\cite{darpa_tc} records system provenance but lacks browser activity and is hard to reproduce.

\textit{Host-based and multi-source datasets:} 
Recent datasets combine host telemetry with network logs for multi-stage attacks~\cite{anjum2021optc, turcotte2018unified}. 
DAPT 2020~\cite{myneni2020dapt}, Unraveled~\cite{myneni2023unraveled}, CICAPT-IIoT~\cite{ghiasvand2024cicaptiiot}, and LMDG~\cite{mabrouk2025lmdg} extend this line but omit browser activity. 
ATLAS~\cite{alsaheel2021atlas} and ATLASv2~\cite{riddle2024atlasv2} are the closest prior work, capturing browser logs alongside system audit and DNS data, but label events only as malicious or benign without \gls{mitre-attack} technique granularity. 
We add per-entry technique labels and broader tactic coverage (12 tactics, 53 techniques).

\textit{Language models for security:} 
Researchers apply language models to malware analysis, vulnerability discovery, log review, and alert triage~\cite{chen2024survey, han2023loggpt}. 
Recent work shows \glspl{llm} identify broad threats effectively but underperform on precise multi-step reasoning~\cite{ullah2024llms, secvuleval2025}. 
Parameter-efficient fine-tuning (e.g., LoRA) lets practitioners adapt models cheaply~\cite{hu2022lora, dettmers2023qlora}.

\textit{Automated \gls{mitre-attack} technique mapping:} 
Prior work links threat-report text to \gls{mitre-attack} techniques~\cite{husari2017ttpdrill, legoy2020rcatt, li2022attackg, alam2023ladder, rani2024ttpxhunter}. 
Unlike these studies, we identify techniques directly from raw, multi-source log data.

No public dataset combines simultaneous system, network, and browser sources with per-entry \gls{mitre-attack} technique labels from real attack tooling. 
Current research also rarely compares base versus fine-tuned \glspl{slm} on multi-source security data. 
We address both gaps with our dataset and a controlled evaluation of locally runnable \glspl{slm} in base and fine-tuned form.

\section{Methodology}
\label{sec:methodology}

We present the methodology below.

\subsection{Dataset Construction}
\label{sec:dataset-construction}

\subsubsection{Overview}
\label{sec:methodology-overview-dataset}

We constructed a cybersecurity log dataset. 
We generated the data through attack simulations where an attacker exploited a victim's system.
To incorporate normal background activity, we executed benign user simulation sessions alongside attack simulation sessions.
To capture a complete picture, we collected information from multiple sources, including system logs, network logs, and browser logs.
To ensure safety, eliminate external variables and keep the dataset reproducible, we controlled the simulation environment and isolated the network. 

To synchronize event timelines, we captured system, network, and browser logs simultaneously.
To manage the synchronized data, we recorded all the sessions in 20-minute blocks.
To replicate real-world scenario, we used genuine attack tools.
Following the methodology from January 2025 to February 2026, we recorded 870 sessions (70 attack simulation sessions and 800 benign user sessions) and generated 2.3 million event logs.

\subsubsection{Environment Setup}

We carried out all our experiments on Windows 10 and Windows 11 systems, using both computers and virtual machines. 
We used computers to simulate physical initial access techniques such as \gls{usb} drive transfers.
We further used \glspl{vm} to scale up our network and simulate a multi-computer environment without needing dozens of physical machines.

We set up the environment to achieve two goals.
First, we had to be able to run live malware (e.g. \gls{rat} and \gls{c2} tunnels) without putting any external system at risk.
Second, we needed to keep things realistic so that the captured network traffic looked exactly like what a security defender would see in real life.
To achieve the goals, we separated the lab from our production networks.
We only connected it to the public internet through a controlled \gls{nat} gateway.
The separation let us limit internet access strictly to the destinations we needed for our \gls{c2} servers, cloud-storage uploads, and tool downloads.
Besides, we used contemporary attack tools instead of synthetic traffic generators to simulate the real-world scenario as detailed in Section~\ref{sec:methodology-overview-dataset}.

Before recording the sessions, we installed our data-collection software stack (detailed in Section~\ref{sec:data-collection}) on every associated system.
Finally, we standardized the setup on all computers to keep the data consistent across different sessions.

\subsubsection{Data Collection}
\label{sec:data-collection}

To coordinate data collection at each monitored endpoint, we developed an orchestrator.
To enable network packet capture, we first verified administrator privileges.
Next, we created a timestamped folder for a session and launched three collection modules (network-capture, system-log, browser-log) as parallel processes.
We monitor these processes for the entire 20-minute session duration.
To save local disk space, we uploaded all artifacts to \texttt{Google Drive} and deleted local copies when a session capture completed.
Finally, to gracefully end interrupted sessions without leaving unfinished tasks behind, we used a cleanup process that automatically closed all active programs.
The three log collection modules are described below in detail.

\paragraph{Network Capture}

To capture all incoming and outgoing network traffic, we invoked \texttt{tshark} in promiscuous mode and without a capture filter.
We saved the observations directly to a \gls{pcap} file.
We did not filter out any network traffic during collection to preserve the full diversity of the protocols for future forensic analysis.

\paragraph{System-Log Capture}

We extracted system logs using an \gls{elk} infrastructure, which is a platform composed of three open-source tools used to collect, parse, and store log data.
During each session, a tracking tool named \gls{sysmon} monitored operating system activity alongside standard \glspl{wel} on the endpoint (i.e. victim machine).
While WEL provides administrative and auditing function, \gls{sysmon} provides low-level, security-specific data.
To keep the data volume manageable while capturing activity, we filter for nine essential \gls{sysmon} event types: process creation, network connection, image and \gls{dll} loading, remote thread creation, file creation, registry-object change, registry-value set, \gls{dns} query, and file deletion.

The pipeline processes the filtered data in a sequence. 
A lightweight shipper program called Winlogbeat continuously reads the local records created by \gls{sysmon} and \gls{wel} and ships them to Logstash. 
Logstash acts as the central data-processing engine that cleans the incoming data, standardizes the formatting, and normalizes the timestamps. 
After parsing, Logstash writes the structured data into Elasticsearch, a database optimized for rapid searching and analytics. 
When the session ends, the orchestrator runs a time-bounded query against Elasticsearch to isolate the exact window of activity and exports these retrieved events into a single \gls{json} document for analysis.

\paragraph{Browser-Log Capture}

We extracted browser logs by querying the local \gls{aw} service on port 5600.
\gls{aw} is an open-source time-tracking application that runs in the computer's operating system and logs user activity.
The service runs two sub-programs: one records the active foreground application, and the other tracks visited \glspl{url} and titles of tab across Chrome, Firefox, and Edge browsers.
We collected all events recorded during each session interval and calculated the duration of each activity within each event.
Then, we sanitized the query parameters to remove \gls{pii} present in the \glspl{url}.
Finally, we export the cleaned activity timeline into a single \gls{json} document.

\subsubsection{Attack Simulation Session}

We conducted controlled attack simulations to generate realistic attack logs by executing sequences of malicious techniques across multiple log sources.

\paragraph{Attack Tool Arsenal}
\label{sec:attack-tools}

We introduce the tools used to achieve our objectives, structured by \gls{mitre-attack} tactic.

\textit{Initial Access / Execution.} 
We used a tool called the \texttt{BD2 .NET Injector} to hide our payloads inside normal files, like images, \glspl{pdf}, and office documents. 
When a victim opens one of these files, they see the normal image or document they expect, while our payload runs silently in the background.

\textit{Defense Evasion.} 
We modified a \gls{rat} called Revenge-\gls{rat} v0.3.
We updated the C\# code of this tool so that when it runs, it automatically turns off Windows Defender's real-time, tamper, and cloud protections, and stops it from sending samples to Microsoft.
Then we signed the modified \gls{rat} with a code-signing certificate to make it look legitimate to security checks.
We also used multiple layers of encryption (e.g. \gls{aes}), encoding (e.g. Base64 encoding), and obfuscation (e.g. IntelliLock) to hide the code and bypass security scanners.

\textit{Privilege Escalation:}
We used \texttt{Process Hacker} to inject our code into trusted Windows processes like \texttt{explorer.exe} and \texttt{svchost.exe}, using process hollowing and \gls{dll} injection.

\textit{Persistence.}
To make sure our access stayed active on the computer, we used three different methods. 
First, we set up scheduled tasks in the \texttt{\gls{wts}} so our code runs automatically on a timer. 
Second, we added entries into the \texttt{Windows Registry} under the standard \texttt{Startup} folders so our code runs whenever someone logs in. 
Finally, we changed the computer's bootloader so our code starts running even before the Windows operating system fully loads.

\textit{Credential Access.}
We retrieved stored credentials by uploading a tool called \texttt{\gls{wbpv}} through our \gls{rat}. 
We then ran that tool using standard command-line programs like \texttt{PowerShell} or \texttt{cmd.exe}.

\textit{Command and Control.}
We sent our \gls{c2} traffic through \texttt{Playit.gg} to bypass network firewalls. 
We configured the tool to use unusual ports (1337 and 4444) and set our software to check in with us every minute.

\textit{Exfiltration.}
We exfiltrated data by uploading it to cloud services. 
We used \texttt{7-Zip} to pack large files into zip archives first, then used \texttt{rclone} to send data to \texttt{Google Drive}, \texttt{mega-cli} to send it to \texttt{Mega.io}, and the built-in \texttt{OneDrive} program to send it to Microsoft cloud storage.

\textit{Impact.}
We simulated a ransomware attack by using \texttt{7-Zip} to encrypt files with a password.
We then deleted the original unencrypted files, and left a ransom note behind.

\paragraph{Attack Scenario Design}

Before executing the attacks, we designed fifty different scenarios to capture a wide array of contemporary hacking methods. 
For each scenario, we selected a unique combination of techniques picking from seven different tactics, as outlined below.

\textit{Initial Access \& Execution.}
We started each attack by selecting an initial access technique, and an execution timing control technique.
As the initial access technique, we selected phishing via Telegram, WhatsApp, email, or fake websites.
Through phishing, we deployed spoofed installer downloads, fake software installers, or hidden payloads inside document or image files.
To determine the execution time of the payloads, we selected either delayed activation, or immediate activation (i.e. \gls{rat} connected instantly).  

\textit{Defense Evasion.}
We used defense evasion techniques during the execution delay stage and at the final stage of the attack lifecycle.
We selected delayed execution to bypass security sandboxes and time-based security controls.
At the end of the operations, we carried out a cleanup activity. 
Either we removed the trace to remove system logs and clear operational histories, or we left some traces behind purposefully for the security teams to identify the attack.

\textit{Credential Access.}
We carried out password harvesting using one of two credential access methods.
We either deployed our \gls{rat} in combination with a extraction utility named \gls{wbpv} to gather saved credentials from browsers, or we relied solely on the \glspl{rat} built-in credential dumping functionalities to extract security tokens from the system.

\textit{Persistence.}
To maintain long-term access to the target system, we established persistence using one of two techniques.
We configured the attack to either create scheduled entries within the \gls{wts} to trigger execution on a specific time, or we modified the system registry, including startup keys and bootloader.

\textit{Lateral Movement.}
We defined the internal scope of each operation by implementing one of three lateral movement techniques.
We either confined the entire attack to a single target system with no internal propagation, expanded our reach across the internal network by abusing Windows \gls{smb} shares, or launched internal spear-phishing campaigns targeting other users inside the same network.

\textit{Exfiltration.}
We moved the collected data out of the target system using one of the nine exfiltration channels.
We routed the stolen information through physical \gls{usb} drive transfers, directed connections to an \gls{ftp} server, custom web uploads, directed transfers through our remote access tool's \gls{c2} server, or hosted cloud storage services including Google Drive, Mega.io, OneDrive, Dropbox, and GitHub .

\textit{Impact.}
We concluded each attack scenario by executing a final payload to achieve one of the three types of operational impact.
We either deployed ransomware to encrypt user files and demanded financial payment, focused strictly on data theft while leaving system functions intact, or executed a system wipeout to erase data and disrupt operations.

\paragraph{Attack Execution Workflow}

We carried out the attacks using a two-person team where one played the attacker and the other played the victim. 
One researcher executed the attack techniques, while the other worked on the victim system to act like a regular user by opening phishing files or running software installers.

We followed a 20-minute timeline for each attack session. 
We triggered the initial entry point right at the start (0:00) and executed our remote-access payload two minutes later (0:02). 
By 0:03, we established our \gls{c2} tunnel through \texttt{Playit.gg}, and at 0:05, we executed reconnaissance commands to look around the system. 
We then targeted and collected saved credentials at 0:06 and installed our persistence mechanisms at 0:08 to maintain access.
If the scenario required it, we moved laterally to other systems at 0:10. 
We began gathering sensitive files at 0:12 and started exfiltrating the collected data out of the network at 0:15. 
Finally, we triggered the impact like running ransomware, stealing data, or wiping the system at 0:18, and wrapped up the entire session by finishing our cleanup phase at 0:20.

We chose the standardized timeline so that every session clearly showed the techniques for each attack step. 
Meanwhile, we had enough spare time in between the steps to adapt to different scenario-specific requirements.
Once the 20-minute window closed, we finalized the session.
We ended the network packet capture, and executed our log extraction modules to pull all system and browser events that happened during the session.

We then uploaded all three files (i.e. packet capture, system log, and browser log) to a timestamped session folder in Google Drive and deleted the temporary copies from the local machine.
Finally, we tagged each session's metadata with its unique scenario ID and labeled it as an attack session.

\subsubsection{Benign User Session}

To identify attacker's behavior and avoid overfitting to attack-only artifacts, detection models require data on benign user behavior.
To accomodate benign user behavior in the dataset, we collected 800 benign user sessions which contained routine system activities.
To capture variability in user behavior, multiple authors carried out the benign user sessions.

We distributed the data across five activity classes: web browsing (35\%), productivity work (25\%), software development (20\%), system administration (10\%), and entertainment (10\%).
The sessions comprise web browsing (280 sessions, including video streaming, information retrieval, news, and social media), productivity work (200 sessions, including office suites and email services), and software development (160 sessions, including code editing, version control, and compilation).
The remaining data consists of system administration (80 sessions, including PowerShell scripting, package management, and maintenance) and entertainment (80 sessions, including gaming and media playback).

We used different procedures for collecting attack and benign sessions.
For the benign session, we carried out tasks without following strict scripts during the 20-minute window.
The resulting data reflects the unpredictable nature of normal computer use, having variations in event timing, and content.
Conversely, for the attack session, we executed the exploits using strict scripts.
The resulting data records the attack scenarios consistently without human error or random noise.

\subsection{Dataset Evaluation}

This section details the dataset evaluation steps.

\subsubsection{Data Preprocessing \& Feature Engineering}
To fine-tune the model, we cleaned the raw logs, added annotations, engineered features, and split the data into train, validation and test sets.

\paragraph{Data Cleaning}

We first converted raw \gls{pcap} into \gls{json} format and parsed the system and browser log files to normalize their timestamps into a single and consistent format.
Next, we secured user privacy by anonymizing \gls{pii}, which involved masking internal \gls{ip} addresses, hashing usernames, and removing sensitive file paths from our logs.
Finally, we mapped all the data fields from our network, system, and browser logs into a single, unified schema to establish relationship between them.

\paragraph{Data Annotation}

To create our ground truth data, we used the attack traces documented in the logs to identify which activities were malicious.
We then labeled every 20-minute log session as either "normal" or "suspicious".
For each "suspicious" session, we identified individual log entries corresponding to malicious activities.
We then mapped the malicious log entries to \gls{mitre-attack} techniques.

\paragraph{Feature Engineering}

To prepare our data, we organized events into chunks (see Section~\ref{subsec:chunk}, which is the fundamental unit of analysis for our \glspl{slm}. 
Empirically, we determined that seven events are sufficient to maintain the necessary context, so each chunk consists of a temporally ordered sequence of seven log entries, accompanied by a metadata object containing the session ID, chunk index, and event count.

Then, we labeled each chunk based on the activities within. 
First, we labeled all the chunks as normal. 
Afterwards, if a chunk contains any malicious events, we labeled it as "suspicious" and mapped each suspicious event to its corresponding \gls{mitre-attack} technique IDs.

Finally, we structured these chunks into an instruction-tuned format to train the models to analyze, reason, and provide evidence. 
The format consists of three fields: an instruction prompt directing the model to analyze the chunk; an input \gls{json} containing the metadata and the ordered sequence of events to facilitate reasoning over attack progression; and an output field providing either a "Normal Activity Detected" confirmation or a structured alert response that includes a severity rating, a list of suspicious events with referenced log fields, and the associated \gls{mitre-attack} technique IDs.

\paragraph{Train-Test-Validation Split}

The feature-engineered dataset comprised $\approx112,726$ chunks, split into a training set ($89,693$), a validation set ($12,427$), and a test set ($10,606$).
To manage GPU memory constraints, we randomly sampled $40\%$ of the training and validation sets, producing $35,877$ and $4,970$ chunks, respectively.
To prevent the truncation of context, we filtered all chunks by a maximum token count of $2000$, resulting in a final training set of $22,896$ chunks and a final validation set of $3,065$ chunks. 

\subsubsection{Model Selection}

We selected three instruction-tuned \glspl{slm} to evaluate the performance $\Delta$ (gain or loss) that occured before and after fine-tuning on our constructed dataset.
To ensure that the performance $\Delta$ generalize broadly rather than being isolated to a specific model ecosystem, we selected \glspl{slm} across three families and parameter sizes.

\begin{table}[htbp]
\vspace{-5pt}
\centering
\scriptsize
\setlength{\tabcolsep}{9pt}
\caption{Overview of Selected \glspl{slm}}
\label{tab:models_overview}
\begin{tabular}{| l | c | c | c |}
\hline
\textbf{\gls{slm} Name} & \textbf{Family} & \textbf{Parameters} & \textbf{Provider} \\ \hline
Qwen2.5-1.5B-Instruct & Qwen-2.5 & 1.5B & Alibaba Cloud \\
Llama-3.2-3B-Instruct & Llama-3.2 & 3.0B & Meta \\
Phi-4-Mini-Instruct & Phi-4 & 3.8B & Microsoft \\ \hline
\end{tabular}
\\[3pt]
\footnotesize \textit{Note:} All listed \glspl{slm} utilize a decoder-only transformer architecture.
\vspace{-5pt}
\end{table}

\subsubsection{Fine-Tuning with \gls{lora}}

We fine-tuned all three \glspl{slm} using \gls{lora}.
The parameter efficient fine tuning technique allows us to train the \glspl{slm} efficiently while keeping the exact same setup across all \glspl{slm} for a fair comparison.

We used a standard configuration for all \glspl{slm}, setting the \glspl{lora} rank $r=16$, scaling factor $\alpha=32$, and a maximum sequence length of $2000$ tokens.
Each \gls{slm} was trained for up to 3 epochs with a batch size of 4, updating less than $1\%$ of the total parameters.

\subsubsection{Evaluation Metrics}

We assessed the performance of \glspl{slm} across ten metrics spanning two primary tasks.

Task 1, chunk classification, determines whether activities within a chunk are normal or suspicious. 
We assessed this performance using seven metrics: accuracy, precision, recall, and F1-scores, reporting both macro-averaged and weighted-averaged figures to account for the data imbalance between normal and suspicious chunks.

Task 2, alert generation, produces the reasoning behind a suspicious classification, the severity level, and the associated \glspl{mitre-attack} techniques for every event. 
We evaluated this task using three metrics: exact-match accuracy, average partial match, and word-level F1 score. 
We used the first two metrics to evaluate whether the generated output matched the reference text perfectly or partially, while the final metric determined the semantic overlap between the generated reasoning and the ground truth.

\section{Findings}
\label{sec:findings}

We present the findings below.

\subsection{Dataset Findings (\texorpdfstring{$RQ_1$}{RQ1})}

We present the findings on the dataset below.

\subsubsection{Dataset Statistics and Quality}

In this section, we report the characteristics of the dataset.

\paragraph{Dataset Composition}

The dataset comprises approximately 870 sessions collected over a 13-month period (January 2025 – February 2026), totalling 290 hours (17,400 minutes) of monitored activity.

\begin{table}[h!]
\vspace{-5pt}
\centering
\small
\caption{Dataset Statistics Overview}
\label{tab:dataset_stats}
\begin{tabular}{| l | l |}
    \toprule
    \textbf{Component} & \textbf{Count / Value} \\
    \midrule
    Total Sessions & $870$ \\
    Attack Sessions & $70$ (8\%) \\
    Normal Sessions & $800$ (92\%) \\
    Session Duration & 20 minutes (fixed) \\
    Total Monitoring Time & 17,400 minutes ($\sim290$ hours) \\
    Total Log Events & $\sim2,300,000$ \\
    Collection Period & January 2025 -- February 2026 \\
    \bottomrule
\end{tabular}
\vspace{-5pt}
\end{table}

The 70 attack sessions were distributed across 8 initial access tactics with Telegram phishing being the most prevalent (17\%), followed by WhatsApp phishing and fake software installers (14\% each).

\paragraph{Log Category Distribution}

Attack sessions produce on average 2.2 times more system log events and 1.9 times more network events than benign sessions.
The difference shows the heavy footprint left behind when our simulated attacks created new processes, modified the system registry, and connected back to \gls{c2} servers.

\begin{table}[h!]
\vspace{-5pt}
\centering
\small
\caption{Log Distribution}
\setlength{\tabcolsep}{10pt}
\label{tab:log_dist}
\begin{tabular}{| l | c | c | c | c | c |}
    \toprule
    \textbf{Type} & \textbf{Total} & \textbf{\%} & \textbf{$\mu$} & \textbf{$\mu_a$} & \textbf{$\mu_n$} \\
    \midrule
    System & 1.8M & 78\% & 2.1k & 4.2k & 1.9k \\
    Network & 400k & 17\% & 460 & 800 & 420 \\
    Browser & 100k & 4\% & 115 & 180 & 110 \\
    \bottomrule
\end{tabular}
\\[5pt]
\footnotesize \textit{Note:} k = thousands; M = millions; $\mu$ = mean entries per session; $\mu_a$ = mean entries per attack session; $\mu_n$ = mean entries per normal session.
\vspace{-5pt}
\end{table}

The most prevalent \gls{sysmon} event type is Process Creation (36\% of system logs), followed by Network Connection (25\%) and Registry Value Set (19\%). 
Attack-specific network indicators include approximately $1,500$ \gls{c2} connections to \texttt{Playit.gg} tunnels on port 1337, $\approx500$ large exfiltration transfers ($>10$ MB), and $\approx800$ DNS queries to suspicious domains.

\paragraph{\gls{mitre-attack} Technique Coverage}

The dataset covers 12 distinct tactics and 53 distinct techniques in the attack sessions (see Table~\ref{tab:mitre_coverage}).

\begin{table}[h!]
\vspace{-5pt}
\centering
\small
\caption{\gls{mitre-attack} Tactic and Session Coverage}
\label{tab:mitre_coverage}
\begin{tabular}{| l | c | c |}
    \toprule
    \textbf{Tactic} & \textbf{Techniques} & \textbf{Coverage (\%)} \\
    \midrule
    Initial Access & 8 & 100\% \\
    Execution & 5 & 100\% \\
    Persistence & 4 & 71\% \\
    Privilege Escalation & 3 & 43\% \\
    Defense Evasion & 7 & 100\% \\
    Credential Access & 2 & 86\% \\
    Discovery & 4 & 100\% \\
    Lateral Movement & 3 & 21\% \\
    Collection & 3 & 100\% \\
    Exfiltration & 9 & 100\% \\
    Command and Control & 2 & 100\% \\
    Impact & 3 & 57\% \\
    \bottomrule
\end{tabular}
\vspace{-5pt}
\end{table}

By achieving 100\% session coverage across seven tactics, our scenarios consistently map out a complete cyber kill chain from delivery to target compromise: Initial Access $\rightarrow$ Execution $\rightarrow$ Defense Evasion $\rightarrow$ Discovery $\rightarrow$ Collection $\rightarrow$ \gls{c2} $\rightarrow$ Exfiltration. 
Lateral Movement has the lowest coverage (21\%) as it was an optional element in the scenario design, included only when the scenario required multi-host propagation.

\paragraph{Dataset Integrity and Validation}

We present the performance across the six dataset integrity dimensions in Table~\ref{tab:quality_metrics}.

\begin{table}[h!]
\vspace{-5pt}
    \centering
    \small
    \caption{Dataset Integrity Metrics}
    \label{tab:quality_metrics}
    \begin{tabular}{| l | c |}
        \toprule
        \textbf{Integrity Metric} & \textbf{Performance} \\
        \midrule
        Session Completeness (all 3 artifacts present) & 100\%       \\
        Packet Capture Success Rate                    & 95\%        \\
        System Log Extraction Success Rate             & 99.5\%      \\
        Browser Log Extraction Success Rate            & 99.8\%      \\
        Timestamp Consistency Across Modalities        & 100\%       \\
        Attack Scenario Compliance                     & 100\%       \\
        \bottomrule
    \end{tabular}
    \vspace{-5pt}
\end{table}

Every session comprises three synchronized artifacts (i.e., \gls{pcap}, system log, and browser log), resulting in a complete multi-source record. 
The extraction rates for system and browser data demonstrate the reliability of the underlying ingestion and logging pipelines, while the matching embedded timestamps confirm temporal alignment across all sources.

The packet capture issue was caused by occasional administrator privilege verification failures at session initialization during promiscuous-mode capture. 
The packet capture drops were isolated and did not corrupt the extraction of the corresponding system and browser logs for those sessions.

\subsection{Dataset Evaluation Findings (\texorpdfstring{$RQ_2$}{RQ2})}

This section reports the outcome of fine-tuning the three \glspl{slm} on our dataset.
The findings are presented in Table~\ref{tab:slm_performance}.

\begin{table}[h!]
\vspace{-5pt}
\centering
\scriptsize
\setlength{\tabcolsep}{7pt}
\caption{Performance metrics comparison}
\label{tab:slm_performance}
\begin{tabular}{|l|cc|cc|cc|}
\hline
\multirow{2}{*}{\textbf{Metric}} & \multicolumn{2}{c}{\textbf{Llama-3.2-3B}} & \multicolumn{2}{c}{\textbf{Phi-4-mini}} & \multicolumn{2}{c}{\textbf{Qwen2.5-1.5B}} \\
\cline{2-7}
 & Base & FT & Base & FT & Base & FT \\
\hline
Accuracy & 0.081 & 0.928 & 0.072 & 0.970 & 0.079 & 0.899 \\
Precision (Macro) & 0.336 & 0.544 & 0.315 & 0.588 & 0.341 & 0.591 \\
Precision (Weighted) & 0.842 & 0.919 & 0.783 & 0.976 & 0.858 & 0.967 \\
Recall (Macro) & 0.296 & 0.459 & 0.254 & 0.646 & 0.270 & 0.587 \\
Recall (Weighted) & 0.081 & 0.928 & 0.072 & 0.970 & 0.079 & 0.899 \\
F1 (Macro) & 0.056 & 0.488 & 0.058 & 0.613 & 0.058 & 0.588 \\
F1 (Weighted) & 0.017 & 0.920 & 0.023 & 0.972 & 0.028 & 0.931 \\
Exact Match Acc. & 0.000 & 0.293 & 0.000 & 0.416 & 0.000 & 0.278 \\
Avg Partial Match & 0.186 & 0.857 & 0.205 & 0.931 & 0.186 & 0.846 \\
Avg F1 (Word-level) & 0.097 & 0.849 & 0.106 & 0.930 & 0.131 & 0.819 \\
\hline
\end{tabular}
\\[3pt]
    \footnotesize \textit{Note:} Base versus fine-tuned performance for all three \glspl{slm} across ten evaluation metrics. FT denotes the LoRA fine-tuned variant.
    \vspace{-5pt}
\end{table}

\subsubsection{Fine-Tuning Gains Performance Across All \glspl{slm}}

Fine-tuning made every model perform better across all ten metrics we tracked. 
Because every result improved and none got worse, we have a clear, positive answer to our second research question.
Before fine-tuning, the base \glspl{slm} had accuracy between 7\% and 8\%.
After fine-tuning, the accuracy improved between 89.9\% and 97.0\%.
Phi-4-mini improved the most, rising to 0.970 ($\Delta=+0.898$).
Llama-3.2-3B improved to 0.928 ($\Delta=0.847$), and Qwen2.5-1.5B improved to 0.899 ($\Delta=0.820$).

\subsubsection{Over-Alerting Behavior in Base \glspl{slm}}

The confusion matrices show that the poor performance of \glspl{slm} was not due to random mistakes, but to one consistent problem: they almost never correctly identified a normal session.  
In fact, the base \glspl{slm} labeled only 0.1\% of normal sessions for Llama-3.2-3B, 0.4\% for Phi-4-mini, and 0.7\% for Qwen2.5-1.5B correctly. 
Instead, they incorrectly marked almost all normal traffic as suspicious. 
Moreover, they frequently assigned an "Unknown" label to 12.3\%, 26.3\%, and 17.7\% of all sessions.

\begin{table}[h!]
\vspace{-5pt}
\centering
\scriptsize
\setlength{\tabcolsep}{4pt}
\caption{Confusion Matrix Comparison (Base vs Fine-Tuned)}
\label{tab:confusion_matrices}
\begin{tabular}{|l|ccc|ccc|ccc|}
\hline
\textbf{} & \multicolumn{3}{c|}{\textbf{Llama-3.2-3B}} & \multicolumn{3}{c|}{\textbf{Phi-4-mini}} & \multicolumn{3}{c}{\textbf{Qwen2.5-1.5B}} \\
\textbf{} & N & S & U & N & S & U & N & S & U \\
\hline
\textbf{Base} & & & & & & & & & \\
Normal (N) & 11 & 8444 & 1193 & 40 & 7047 & 2561 & 71 & 7879 & 1698 \\
Suspicious (S) & 1 & 849 & 108 & 7 & 726 & 225 & 5 & 770 & 183 \\
Unknown (U) & 0 & 0 & 0 & 0 & 0 & 0 & 0 & 0 & 0 \\
\hline
\textbf{Fine-Tuned} & & & & & & & & & \\
Normal (N) & 9467 & 171 & 10 & 9356 & 281 & 11 & 8717 & 220 & 711 \\
Suspicious (S) & 580 & 378 & 0 & 29 & 929 & 0 & 136 & 822 & 0 \\
Unknown (U) & 0 & 0 & 0 & 0 & 0 & 0 & 0 & 0 & 0 \\
\hline
\end{tabular}
\\[5pt]
    \footnotesize \textit{Note:} Rows represent true labels, and columns represent predicted labels.
    \vspace{-5pt}
\end{table}

The base \glspl{slm} seem to detect attacks well, but only because they indiscriminately label nearly everything as suspicious. 
This behavior flags almost all benign traffic as threats. 
The gap between weighted and macro scores confirms the finding: because normal sessions dominate the data, mislabeling them forces accuracy and weighted F1 towards zero, despite decent macro precision of 0.32-0.34. 
Consequently, the base \glspl{slm} are operationally unusable as they would overwhelm analysts with false positives.

\subsubsection{Fine-Tuning Fixes Over-Alerting, but \glspl{slm} Performance Varies in Attack Detection}

Fine-tuning fixed the low-recall problem in every \gls{slm}. 
The \glspl{slm} started correctly identifying normal sessions with recall improvement of 98.0\% for Llama-3.2-3B, 96.6\% for Phi-4-mini, and 89.6\% for Qwen2.5-1.5B. 
Additionally, the \glspl{slm} almost entirely stopped making "Unknown" predictions, which had previously accounted for up to 26.3\% of outputs in the base variants.

However, the \glspl{slm} show variation when detecting the "Suspicious" sessions, which are rare.
Phi-4-mini performed the best, correctly identifying 97.0\% of suspicious sessions.
Qwen2.5-1.5B followed with 85.8\%.
Llama-3.2-3B underperformed, identifying only 39.5\% of the "Suspicious" sessions and misclassifying 580 out of 958 attacks as "Normal".
 
Phi-4-mini, the largest fine-tuned \gls{slm} at 3.8B parameters, remains the most reliable, outperforming the others in all ten metrics.

\subsubsection{Limitation in Technique Identification Task}

None of the base \glspl{slm} could produce an exact correct response with \gls{mitre-attack} technique and response, resulting in an exact-match accuracy of 0 across all three \glspl{slm}.
After fine-tuning, Phi-4-mini achieved an exact-match accuracy of 0.416, followed by Llama-3.2-3B at 0.293, and Qwen2.5-1.5B at 0.278.
Besides, higher partial-match scores (0.857-0.931) and word-level F1 scores (0.819-0.930) indicate that fine-tuned \glspl{slm} identified most correct techniques and produced more relevant reasoning even when failing to reproduce the exact output.

\section{Discussion}
\label{discussion}

The fact that fine-tuning consistently improved three different \gls{slm} from near-zero to 90–97\% accuracy indicates that our dataset is learnable and contains consistent signals that various \gls{slm} family can extract. 
When combined with the dataset’s broad coverage (12 tactics and 53 techniques), the finding shows that controlled simulations can create high-quality, multi-source log datasets suitable for training \glspl{slm}.

For security applications, both macro and weighted metrics are worthy of consideration: macro metrics assess the \glspl{slm}'s ability to detect rare threats by treating all classes equally, while weighted metrics evaluate overall detection capability across the entire dataset. 
By these measures, Phi-4-mini is the most effective \gls{slm}, leading in both macro and weighted categories. 
While the largest \gls{slm} (Phi-4-mini) performed best, the sample size is insufficient to conclude that larger fine-tuned models consistently generalize better.

Because base \glspl{slm} trigger a flood of false alarms, they are operationally unusable.
Fine-tuned \glspl{slm} are reliable enough to perform the initial triage.
The high partial-match and word-level scores suggest that fine-tuned \glspl{slm} can support human analysts by suggesting relevant techniques and reasoning.

\section{Threats to Validity}
\label{sec:threats-to-validity}

\subsubsection{Internal Validity}

\glspl{slm} might be identifying patterns in the simulation's structure or output formatting rather than actual malicious behavior, which could artificially inflate performance results. 
Additionally, using a uniform LoRA configuration ensures a fair comparison between models, even if it does not represent the absolute maximum potential of each individual model.

\subsubsection{External Validity}

As the dataset was created in a controlled lab environment, the accuracy levels may not fully translate to the more diverse conditions of enterprise networks. 
Additionally, factors like the specific percentage of attacks and the limited variety of user behavior in the test data could cause performance results to shift in a production environment.

\subsubsection{Construct Validity}

As accuracy is misleading with imbalanced data, we used macro-averaged metrics and per-class recall to provide a more accurate picture of model performance. 
We also combined strict exact-match scores with partial-match and word-level F1 to ensure we do not undercount correct identifications that differ in format.
The team that designed the attacks assigned the ground-truth labels, which introduced a potential annotation bias.

\subsubsection{Conclusion Validity}

As we only tested three \glspl{slm} once, our results indicate trends rather than proven rankings. 
While the massive performance jump from base \glspl{slm} to fine-tuned \glspl{slm} is genuine, the small differences in ranking between fine-tuned \glspl{slm} should be seen as observations rather than concrete proof of superiority.

\section{Conclusion}
\label{sec:conclusion}

We built a multi-source cybersecurity log dataset from controlled attack simulations. 
We then tested three \glspl{slm} before and after fine-tuning on our dataset. 
Fine-tuning produced large and consistent gains. 
Accuracy rose from about 8\% to between 90\% and 97\%. 
Base models failed by marking almost every session as suspicious, so they are not usable in practice. 
Fine-tuning fixed the failure, but the models still differ on rare attacks. 
Phi-4-mini caught 97\% of attacks, while Llama-3.2-3B caught under 40\%.
Technique identification remains difficult, with the best exact match at 42\%. 
Future work should explore fine-tuning a broader range of \glspl{slm}, evaluate performance on production logs, conduct repeated trials, and develop improved methods to close the technique identification gap.

\bibliographystyle{IEEEtran}
\bibliography{references}

\end{document}